\input psfig
\documentstyle[preprint,aps]{revtex}
\draft

\begin{document}
\title{
Monte Carlo Comparison of Quasielectron Wave Functions}
\author{V. Melik-Alaverdian and N.\ E. Bonesteel}
\address{
National High Magnetic Field Laboratory and Department of Physics,
Florida State University, Tallahassee, FL 32306-4005}
\maketitle
\begin{abstract}
Variational Monte Carlo calculations of the quasielectron and
quasihole excitation energies in the fractional quantum Hall effect
have been carried out at filling fractions $\nu=1/3$, 1/5, and 1/7.
For the quasielectron both the trial wave function originally proposed
by Laughlin and the composite fermion wave function proposed by Jain
have been used.  We find that for long-range Coulomb interactions the
results obtained using these two wave functions are essentially the
same, though the energy gap obtained using the composite fermion
quasielectron is slightly smaller, and closer to extrapolated
exact-diagonalization results.
\end{abstract}


\section{Introduction}

Shortly after the discovery of the fractional quantum Hall effect
(fractional QHE) \cite{tsui} Laughlin \cite{laughlin} introduced a set
of trial wave functions describing the `parent' quantum Hall states
occurring at Landau level filling fraction $\nu=1/q$ where $q$ is an
odd integer.  In addition to the ground state, Laughlin introduced
trial wave functions describing fractionally charged quasielectron
$(e/q)$ and quasihole $(-e/q)$ excitations.  From the very beginning
it was clear that the wave function for the quasihole, with its simple
Jastrow form, was more natural than the wave function for the
quasielectron, which contains explicit derivatives with respect to
electron coordinates.  This difference is reflected, for example, in
the fact that while there exists a Hamiltonian for which the Laughlin
ground state and quasihole wave functions are exact (zero energy)
eigenstates\cite{haldane} there exists no such simple Hamiltonian for
which the quasielectron wave function is also an exact eigenstate.

According to the composite fermion theory, proposed by Jain
\cite{jain}, the fractional QHE corresponds to an {\it integer} QHE
of composite fermions -- electrons bound to an even number of
statistical flux quanta.  This identification leads to a procedure for
constructing fractional QHE trial states by first constructing integer
QHE states, then multiplying by a Jastrow factor which binds an even
number of vortices to each electron, and finally projecting the
resulting state onto the lowest Landau level.  The wave function
obtained using this procedure for the parent quantum Hall state is
identical to Laughlin's ground state, and the same is true for the
quasihole wave function.  However, the composite fermion quasielectron
wave function is not identical to the one proposed by Laughlin.

To date, the best estimate of the energy gap for creating a
quasielectron-quasihole pair with infinite separation at $\nu=1/3$
computed using Laughlin's trial states was obtained by Morf and
Halperin, using the disk\cite{morfd} and spherical\cite{morfs}
geometries, with the result $\Delta_L \simeq 0.092 \pm 0.004
e^2/\epsilon l_0$ where $l_0 = \sqrt{\hbar c /eB}$ is the magnetic
length.  This may be compared with the result of Bonesteel
\cite{bonesteel} using the composite fermion quasielectron wave
function of $\Delta_{CF} \simeq 0.106 \pm 0.003 e^2/\epsilon l_0$.  It
therefore appears that the energy gap computed using the Laughlin
quasielectron is over \%10 lower than that obtained using the
composite fermion theory, and so one might conclude that the Laughlin
quasielectron wave function provides a (slightly) better description
of the physical quasielectron state.  More recently, Girlich and
Hellmund\cite{girlich} have shown that for the truncated ($V_1$)
pseudopotential interaction introduced by Haldane \cite{haldane}, the
interaction for which the Laughlin ground state and quasihole state
are exact eigenstates, the Laughlin quasielectron has an energy which
is 18 \% {\it higher} than the composite fermion quasielectron. These
authors go on to speculate that the same would be true for the Coulomb
interaction, and this has motivated us to reexamine the calculations
of Morf and Halperin.

Taking advantage of the availability of significantly faster
computers, and performing a better extrapolation of finite size
results to the thermodynamic limit, we find that the $\nu=1/3$ energy
gap computed using the Laughlin quasielectron extrapolates to
$\Delta_L \simeq 0.110(2) e^2/\epsilon l_0$, a significantly higher
result than previously reported.  This result is higher than
$\Delta_{CF}$ and so is consistent with Girlich and Hellmund
\cite{girlich}, though we find that for the Coulomb interaction the
difference between the two energy gaps is quite small (less than  5 \%)
indicating that both wave functions provide adequate descriptions of
the true quasielectron.  We have also performed what we believe are
the first calculations of the energy gap using Laughlin's
quasielectron wave functions for filling fractions $\nu=1/5$ and 1/7.
Comparing these energies to the corresponding composite fermion
energies we find the same result --- the composite fermion energy gap
is consistently smaller than the corresponding Laughlin energy gap.
Comparing these results to the extrapolated exact diagonalization
results of Fano et al. \cite{fano} we find that the composite fermion
energy gaps are also consistently closer to the `exact' results.
However, as for $\nu=1/3$, the differences are slight, and the main
conclusion is that both wave functions provide an adequate description
of the physical quasielectron.

This paper is organized as follows.  In Sec.~II we review the
formulation of the two dimensional electron gas on the Haldane sphere
and introduce a procedure for projecting wave functions on the sphere
into the lowest Landau level.  In Sec.~III we review both the
composite fermion construction of the quasielectron state, which
requires the projection developed in Sec.~II, as well as the Laughlin
quasielectron wave function.  Finally, Sec.~IV contains a summary of
our results.

\section{The Two Dimensional Electron Gas on the Haldane Sphere}

\subsection{One Particle and Landau Level Projection}

We begin by reviewing the problem of one spin polarized electron
confined to move on the surface of a sphere of radius $R$ with a
magnetic monopole at its center.  Following convention we let
\begin{eqnarray}
2S = \frac{4\pi R^2 B}{hc/e} = 2 \left(\frac{R}{l_0}\right)^2
\nonumber
\end{eqnarray}
denote the number of flux quanta piercing the surface of the sphere
due to the monopole.  The Hamiltonian describing this particle is then
\begin{eqnarray}
T = \frac{|{\bf \Lambda}|^2}{2m R^2}
\end{eqnarray}
where ${\bf \Lambda} = {\bf r} \times ( - i \hbar {\bf \nabla} + e
{\bf A}( {\bf r}))$ and $\nabla \times {\bf A} = B {\bf \hat r}$ on
the surface of the sphere.  In what follows we work in the Wu-Yang
gauge \cite{wu} for which
\begin{eqnarray}
{\bf A} = \frac{\hbar S}{e R} \frac{1-\cos\theta}{\sin\theta} {\bf
e_\phi}
\end{eqnarray}
and use the complex stereographic coordinate $z = x+iy = \cot
\frac{\theta}{2} e^{i\phi}$.   The eigenfunctions of $T$ in
this gauge are the monopole harmonics\cite{wu}
\begin{eqnarray}
Y_{Slm} = M_{Slm}
\left(
\frac{1}{1+|z|^2}\right)^S
z^{S+m} P_{l-s}^{S+m,S-m} 
\left(\frac{1-|z|^2}{1+|z|^2}\right)
\end{eqnarray}
where
\begin{eqnarray}
M_{Slm} = \left(\frac{2l+1}{4\pi} 
\frac{(l-s)! (l+s)!}{(l-m)! (l+m)!}\right)^{1/2},
\end{eqnarray}
$P_n^{\alpha,\beta}$ is a Jacobi polynomial, $l = S,S+1,S+2,\cdots$,
and for a given $l$, $m = -l,-l+1,\cdots,l-1,l$. If we let $n = S-l$
then the energies of these states are
\begin{eqnarray}
E_n = \hbar\omega\left( n+\frac{1}{2} + \frac{n(n+1)}{2S}\right),
\ \ \ \ \ n=0,1,2,\cdots
\end{eqnarray} 
and $n$ is the spherical Landau level index.

It will be necessary in what follows to project general wave functions
onto the lowest Landau level $(n=0)$ Hilbert space.  Following Girvin
and Jach \cite{girvin} we now introduce a general procedure for
performing such a projection on the sphere.  First note that the
lowest Landau level wave functions in the Wu-Yang gauge are
\begin{eqnarray}
\psi_m = \left(\frac{2S+1}{4\pi} 
\left(
\begin{array}{c} 2S \\ S+M \end{array}
\right)\right)^{1/2}
\left(\frac{1}{1+|z|^2}\right)^S z^p,
\ \ \ \ 
p=0,1,\cdots,2S+1.
\nonumber
\end{eqnarray}
The Hilbert space of lowest Landau level wave functions on the sphere
then corresponds to wave functions of the form
\begin{eqnarray}
\psi(z,z^*) = \left(\frac{1}{1+|z|^2}\right)^S\ f(z)
\label{wfs}
\end{eqnarray}
where $f(z)$ is a polynomial of degree up to $2S+1$ in $z$.

The differential area element on the surface of the sphere in terms of
the stereographic coordinates $x$ and $y$ is
\begin{eqnarray}
dA = \frac{4R^2}{(1+|z|^2)^2} dx dy
\end{eqnarray}
and so the scalar product between any two polynomials $f$ and $g$ in
this Hilbert space is
\begin{eqnarray}
\langle f,g \rangle = \int  \frac{4 R^2}{({1+|z|^2})^{2S+2}} f^* g\  dx~dy.
\end{eqnarray}
With this definition of the scalar product it is straightforward to
derive the following identity by repeatedly integrating by parts,
exploiting the fact that $\frac{d}{dz} (f(z)^*) = 0$,
\begin{eqnarray}
\left\langle f, \frac{d^n}{dz^n} g \right\rangle
 = \frac{(2S+2)!}{(2S+2-n)!} \left\langle f, 
\left(\frac{z^*}{1+|z|^2}\right)^n g \right\rangle.
\end{eqnarray}
This result immediately implies the following spherical generalization
of the $z^* \rightarrow d/dz$ rule of Girvin and Jach to the sphere,
\begin{eqnarray}
\left(\frac{z^*}{1+|z|^2}\right)^n \Rightarrow
\frac{(2S+2-n)!}{(2S+2)!} \frac{d^n}{dz^n}
\label{identity}
\end{eqnarray}
Thus, to project any polynomial $f(z,z^*/(1+|z|^2))$ into the lowest
Landau level one simply orders each term so that the $z^*/(1+|z|^2)$'s
all sit to the left, then replaces these factors by derivatives with
respect to $z$ according to (\ref{identity}).

\subsection{$N$ particles and the fractional QHE}

The Hamiltonian for $N$ spin polarized electrons on the sphere
interacting via the Coulomb repulsion is then
\begin{eqnarray}
H = \sum_{i=1}^N T_i + V.
\end{eqnarray}
The interaction energy $V$ is
\begin{eqnarray}
V = \sum_{i<j} \frac{e^2}{r_{ij}} + \frac{1}{2} \frac{Q^2}{R} -
\frac{NeQ}{R},
\end{eqnarray}
where $r_{ij}$ is the chord distance between a given pair of electrons
on the sphere. Here the sphere is assumed to have a uniform
compensating charge density with total charge $Q$.  When considering a
homogeneous state the appropriate background charge is $Q = Ne$ for
which
\begin{eqnarray}
V = \sum_{i<j} \frac{e^2}{r_{ij}} - \frac{1}{2} \frac{N^2e^2}{R}.
\end{eqnarray}

The spherical analog of the Laughlin state\cite{haldane} at $\nu=1/q$
occurs when $q(N-1) = 2S$ and in the Wu-Yang gauge is
\begin{eqnarray}
\psi_{gs} = \prod_k \left(\frac{1}{1+|z_k|^2}\right)^S \prod_{i < j } (z_i - z_j)^q.
\label{laughlings}
\end{eqnarray}
The quasihole wave function corresponding to a single charge $-e/q$
defect at the top of the sphere ($z=0$) is
\begin{eqnarray}
\psi_{qh} = \prod_k \left(\frac{1}{1+|z_k|^2}\right)^S z_k \prod_{i < j } (z_i - z_j)^q.
\label{laughlinqh}
\end{eqnarray}

\section{Quasielectron Wave Functions}

\subsection{Composite Fermion Quasielectron Wave Function}

According to Jain's composite fermion approach \cite{jain} a given
fractional QHE wave function at filling fraction $\nu = p/(kp+1)$
where $k=2,4,\cdots$ is found by first constructing the corresponding
{\it integer} QHE wave function at $\nu_{CF} = p$ and then multiplying
by a Jastrow factor which ties $k$ vortices to each electron.  The
state is then explicitly projected into the lowest Landau level.
Denoting the Slater determinant corresponding to the effective integer
QHE state, with the overall $\prod(1+|z|^2)^{-\tilde S}$ factor
removed, as $\Phi_{CF}$, the corresponding fractional QHE states are
\begin{eqnarray}
\psi = P_{\rm LLL} \prod \left(\frac{1}{1+|z|^2}\right)^S \prod_{i<j} (z_i - z_j)^n
\Phi_{CF}
\label{cfwavefunction}
\end{eqnarray}
where $P_{\rm LLL}$ is the projection operator onto the lowest Landau
level.

For the $\nu_{CF} = 1$ ground state the Vandermonde determinant
\begin{eqnarray}
\Phi_{CF} = \left |
\begin{array}{ccccc}
1      & z_1 & z_1^2 & \cdots & z_1^{N-1} \\
\vdots &     &       &        & \vdots \\
1      & z_N & z_N^2 & \cdots & z_N^{N-1}
\end{array}
\right |
= \prod_{i<j}(z_i - z_j)
\end{eqnarray}
corresponding to one filled `pseudo'-Landau level of composite
fermions, gives for $\psi$ the Laughlin wave function
(\ref{laughlings}) for $\nu=1/q$ where $q = k+1$.  If we remove a
composite fermion from the lowest `pseudo'-Landau level then
\begin{eqnarray}
\Phi_{CF} = \left |
\begin{array}{cccc}
z_1     & z_1^2 & \cdots & z_1^{N-1} \\
\vdots &     &           & \vdots \\
z_N     & z_N^2 & \cdots & z_N^{N-1}
\end{array}
\right |
= \prod_k {z_k} \prod_{i<j}(z_i - z_j) 
\end{eqnarray}
and $\psi$ is identical to (\ref{laughlinqh}) and describes a state in
which a single quasihole sits at the top of the sphere.

We now consider the quasielectron wave function constructed using this
approach.  If we introduce a composite fermion into the first excited
`pseudo'-Landau level then
\begin{eqnarray}
\phi_{CF} = 
\left|
\begin{array}{ccccc}
1 & z_1 & ... & z_1^{N-2} & \frac{{z_1^*}}{1+|z_1|^2} \\
\vdots & \vdots  &  &\vdots &\vdots  \\
1 & z_N & ... & z_N^{N-2} & \frac{z_N^*}{1+|z_N|^2} \\
\end{array}\right|.
\end{eqnarray}
and the corresponding physical electron wave function is
\begin{eqnarray}
\psi_{qe}^{CF} = P_{\rm LLL} \prod_k
\left(\frac{1}{1+|z_k|^{2}}\right)^S
\prod_{i<j}(z_i-z_j)^{q-1} \left|   
\begin{array}{ccccc}
1 & z_1 & ... & z_1^{N-2} & \frac{{z_1^*}}{1+|z_1|^2} \\
\vdots & \vdots  &  &\vdots &\vdots  \\
1 & z_N & ... & z_N^{N-2} & \frac{{z_N^*}}{1+|z_N|^2} \\
\end{array}\right|.
\end{eqnarray}

This wave function can be projected into the lowest Landau level as
follows.  First rewrite $\psi^{CF}_{qh}$ by pulling the Jastrow factor
and the projection operator into the last column of the determinant,
\begin{eqnarray} \psi^{CF}_{qh} = \prod_k
\left(\frac{1}{1+|z_k|^{2}}\right)^{S} \left| \begin{array}{ccccc} 1 &
z_1 & ... & z_1^{N-2} & P_{\rm LLL} \frac{{ z_1^*}}{1+|z_1|^2}
\prod_{i<j}(z_i-z_j)^{q-1} \\ \vdots & \vdots & &\vdots &\vdots \\ 1 &
z_N & ... & z_N^{N-2} & P_{\rm LLL}\frac{{
z_N^*}}{1+|z_N|^2}\prod_{i<j}(z_i-z_j)^{q-1} \\ \end{array}\right|.
\end{eqnarray} This can be done here because the cofactor associated
with the $n^{th}$ element of the $N^{th}$ column does not contain
$z_n$.  Thus, when we do the projection we need only project each
element of the matrix separately.  Following the procedure outlined in
Sec.~IIA this projection gives \begin{eqnarray} P_{\rm
LLL}\frac{{ z_1^*}}{1+|z_1|^2}\prod_{i<j}(z_i-z_j)^{q-1} =
\frac{1}{2S+2} \frac{\partial}{\partial z_1}
\prod_{i<j}(z_i-z_j)^{q-1} =
\frac{q-1}{2S+2}\prod_{i<j}(z_i-z_j)^{q-1} \sum_{i\ne 1}
\frac{1}{z_1-z_i}.  \end{eqnarray} Thus $\psi^{CF}_{qe}$ can be
rewritten, up to an irrelevant normalization constant, to give
\begin{eqnarray} \psi^{CF}_{qe} = \prod_k
\left(\frac{1}{1+|z_k|^{2}}\right)^{S} \prod_{i<j}(z_i-z_j)^{q-1}
\left| \begin{array}{ccccc} 1 & z_1 & ... & z_1^{N-2} & \sum_{i\ne 1}
\frac{1}{z_1-z_i} \\ \vdots & \vdots & &\vdots &\vdots \\

1 & z_N & ... & z_N^{N-2} & \sum_{i\ne N} \frac{1}{z_N-z_i} \\
\end{array}\right|.
\end{eqnarray}

One can simplify things further by expanding the determinant in a
cofactor expansion down the $N^{th}$ column.  The cofactors are then
all Vandermonde determinants and the final expression for the Jain
quasielectron is
\begin{eqnarray}
\psi^{CF}_{qe} = \sum_n \prod_{k \ne n} \frac{1}{z_k - z_n} \sum_{l \ne n}
\frac{1}{z_l - z_n} \psi_{\rm gs}.
\end{eqnarray}
In this form $|\psi|^2$ can be sampled by usual variational Monte
Carlo techniques with each Monte Carlo step taking order $N$
instructions, rather than $N^2$ for a usual determinant. 

\subsection{Laughlin Quasielectron Wave Function}

The generalization to the spherical geometry of the quasielectron wave
function introduced by Laughlin is
\begin{eqnarray}
\psi^{L}_{qe} = 
\left(\prod_k
\left(\frac{1}{1+|z_k|^{2}}\right)^{S}
\frac{\partial}{\partial z_k}
\right)
\prod_{i<j}(z_i-z_j)^q 
\end{eqnarray}
Straightforward Monte Carlo sampling of $|\psi_{qp}|^2$ is not
possible because of the explicit derivatives with respect to the
electron coordinates.  To compute the energy of this state we
therefore follow the procedure of Morf and Halperin\cite{morfd,morfs}
which, for completeness, we review below.  A more detailed discussion
can be found in \cite{morfs}.

Following \cite{laughlin} we first take the absolute square of the
wave function to obtain
\begin{eqnarray}
|\psi^L_{qe}|^2 &=&
\left(\prod_k
\left(\frac{1}{1+|z_k|^{2}}\right)^{2S}
\frac{\partial}{\partial z_k}
\frac{\partial}{\partial z_k^*}
\right)
\prod_{i<j}|z_i-z_j|^{2q} \nonumber \\ \nonumber \\
&=&
\left(
\prod_k 
\left(\frac{1}{1+|z_k|^{2}}\right)^{2S} \frac{1}{4} {\bf \nabla}_k^2 
\right)
\prod_{i<j}|z_i-z_j|^{2q}. 
\end{eqnarray}
The expectation value of any operator $O$ depending only on the
coordinates $\{(x_i,y_i)\}$ is then
\begin{eqnarray}
\langle O \rangle = 
\frac{
\int \left(\frac{1}{1+|z_k|^{2}}\right)^{2S+2} O
{\bf \nabla}_k^2
\prod_{i<j}|z_i-z_j|^{2q} \prod_i dx_i dy_i}
{\int \left(\frac{1}{1+|z_k|^{2}}\right)^{2S+2}
{\bf \nabla}_k^2 
\prod_{i<j}|z_i-z_j|^{2q} \prod_i dx_i dy_i}
\end{eqnarray}
which, after integrating by parts twice in the numerator and the
denominator can be rewritten
\begin{eqnarray}
\langle O \rangle = 
\frac{\int P(z_1,...,z_N) \tilde O \prod_i dx_i dy_i}
{\int P(z_1,...,z_N) \prod_i dx_i dy_i}
\label{exp}
\end{eqnarray}
where
\begin{eqnarray}
P = \left(\prod_k 
\left(
\frac{1}{1+|z_i|^2}
\right)^{2S+4}
\left(
|z_k|^2 - \frac{1}{2S+2} 
\right)\right)
\prod_{i<j} |z_i - z_j|^{2q}
\end{eqnarray}
and
\begin{eqnarray}
\tilde O
= \frac{\prod_j \nabla_j^2 \left(\frac{1}{1+|z_j|^2}\right)^{2S+4} O}
{\prod_j \nabla_j^2 \left(\frac{1}{1+|z_j|^2}\right)^{2S+4}}
\end{eqnarray}
The chord distance between any two points on the sphere is given by
$r_{ij} = R|z_i-z_j|/\sqrt{(1+|z_i|^2)(1+|z_j|^2)}$ and so for the
Coulomb interaction the operator $O$ is
\begin{eqnarray}
V_{\rm Coul.} = \frac{e^2}{R} 
\frac{\sqrt{1+|z_i|^2}\sqrt{1+|z_j|^2}}
{|z_i - z_j|}
\end{eqnarray}
It is then straightforward to compute $\tilde O$ and evaluate
(\ref{exp}) by usual variational Monte Carlo techniques.

\section{Results}

The excitation energies of isolated quasielectron and quasihole states
have been obtained using the trial wave functions reviewed in Secs. II
and III.  Following \cite{morfs} we have computed the {\it proper}
energies, meaning that the relevant ground state energies are computed
with monopole strength $2S = q(N-1)$ and background charge $Q = Ne$,
while the energy of the quasielectron (quasihole) excitations are
obtained keeping $R$ and $N$ fixed and decreasing (increasing) the
monopole strength according to $S \rightarrow S - 1/2$ ($S \rightarrow
S+1/2$).  In addition, following \cite{haldane} and \cite{fano}, we
have shifted the background charge when computing the quasielectron
(quasihole) energies taking $Q\rightarrow Q -e/q$ ($Q \rightarrow Q
+e/q$), in order to compensate the charge density of the {\it bulk} of
the wave function.  This eliminates a finite size correction of $\pm
(e/q)^2 1/R \sim O(\frac{1}{\sqrt{N}})$, a correction which was not
included in \cite{morfs} and which may account for the slightly
different results obtained here.  Our results for the proper energies
of the quasiholes, the Laughlin quasielectrons, and the composite
fermion quasielectrons for filling fractions $\nu=1/3,1/5$ and 1/7 are
shown plotted vs. $1/N$ in Fig.~1.

The quasihole energy gaps obtained here, $\Delta^{q.h.}$, extrapolated
to the $N \rightarrow \infty$ limit are given in Table I, together
with the $\nu=1/3$ result of Morf and Halperin \cite{morfs} and the
extrapolated exact diagonalization results of Fano et al. \cite{fano}.
As stated above, the discrepancy between our results and those of Morf
and Halperin is most likely due to the $1/\sqrt{N}$ finite size
correction we have eliminated by modifying the background charge
density.

The quasielectron energy gaps obtained using both the Laughlin trial
state, $\Delta^{q.e.}_L$, and the composite fermion state,
$\Delta^{q.e.}_{CF}$ are givin in Table II.  Again, in comparing the
present $\nu=1/3$ result for $\Delta^{q.e.}_L$ with those of
\cite{morfs} we note a slight discrepancy which we attribute to the
$1/\sqrt{N}$ finite size correction we have included here.  For
$\nu=1/3,1/5,$ and 1/7 the composite fermion excitation energy
$\Delta^{q.e.}_{CF}$ is consistently about 10 \% lower than the
Laughlin excitation energy, as can be seen clearly in Fig.~1. Note
that the composite fermion result is also in slightly better agreement
with the extrapolated exact diagonalization results of Fano et al.
\cite{fano}.

Finally, Table III gives results for the total energy gap for creating
a well separated quasielectron quasihole pair, $\Delta = \Delta^{q.h.}
+ \Delta^{q.e.}$.  The results are again compared with those of
\cite{morfs} for the Laughlin energy gap as well as those of
\cite{bonesteel} for the composite fermion energy gap. For $\nu=1/3$
our extrapolated energy gap using the Laughlin quasielectron is
$\Delta_L = 0.110 \pm 0.002$, roughly 20 \% larger than the earlier
estimate of Morf and Halperin \cite{morfs}.  Our improved calculation
gives a Laughlin energy gap which is slightly larger than the
corresponding energy gap computed using the composite fermion
quasielectron, $\Delta_{CF} = 0.106 \pm 0.002$.  This is consistent
with the results of Girlich and Hellmund \cite{girlich} using the
short-range $V_1$ model; however, we find here that for the Coulomb
interaction the energy gaps obtained using these two wave functions
are essentially the same.  Table III gives similar results for
$\nu=1/5$ and 1/7.  We therefore conclude that that both the Laughlin
and composite fermion quasielectron wave functions provide adequate
descriptions of the physical quasielectron, though the energy gap
obtained using the composite fermion quasielectron is slightly closer
to extrapolated exact-diagonalization results for all the filling
factors we have considered.

To summarize, the quasielectron and quasihole excitation energies in
the fractional QHE have been calculated for $\nu=1/3,1/5$ and 1/7 by
variational Monte Carlo.  Results have been obtained using the
quasielectron states originally proposed by Laughlin, as well as the
fully projected composite fermion quasielectron states proposed by
Jain.  We have improved on earlier estimates \cite{morfd,morfs} of the
excitation energies of the Laughlin states at $\nu=1/3$ in order to
show that the composite-fermion energy gap is actually slightly lower
than the Laughlin energy gap, consistent with the results of Girlich
and Hellmund \cite{girlich}.  Results for the energy gap using
Laughlin's quasielectron for $\nu=1/5$ and 1/7, obtained here for the
first time, show that, as for $\nu=1/3$, for Coulomb interactions the
energy gaps obtained using the Laughlin and composite fermion
quasielectron wave functions are essentially the same, though those
obtained using the composite fermion quasielectrons are slightly
smaller, and closer to extrapolated exact-diagonalization results of
Fano et al. \cite{fano}, than those obtained using the Laughlin
quasielectron.

\acknowledgements

The authors would like to thank J. Jain and R. Morf for useful
discussions.  This work was supported in part by DOE grant No.\
DE-FG02-97ER45639 and by the National High Magnetic Field Laboratory
at Florida State University.  NEB acknowledges support from the Alfred
P. Sloan Foundation.

\begin{table}
\caption{
Quasihole energy for the fractional QHE with $\nu = 1/3$, 1/5, and
1/7. The Monte Carlo results of Morf and Halperin \protect\cite{morfs}
and extrapolated exact diagonalization results of Fano et al.
\protect\cite{fano} are given for comparison.}
\begin{tabular}{ccccc}
& $\nu$ & $\Delta^{\rm q.h.}$ & $\Delta^{\rm q.h.} (Ref.[1])$ &
$\Delta^{\rm q.h .} (Ref.[2])$ \\
\tableline
& 1/3     &  0.0279(12)        & 0.0224(16)    & 0.0264 \\
& 1/5     &  0.0092(6)         &  ---          & 0.0071 \\
& 1/7     &  0.0038(4)         &  ---          &   ---  \\
\end{tabular}
\end{table}

\begin{table}
\caption{
Quasielectron energy for the fractional QHE with $\nu = 1/3$, 1/5, and
1/7. The Monte Carlo results of Morf and Halperin \protect\cite{morfs}
and extrapolated exact diagonalization results of Fano {\it et al.}
\protect\cite{fano} are given for comparison.}
\begin{tabular}{cccccc}
& $\nu$ & $\Delta^{\rm q.e.}_{\rm CF}$ & $\Delta^{\rm q.e.}_{\rm L}$ &
$\Delta^{\rm q.e.}_{\rm L} (Ref.\protect\cite{morfs})$ & $\Delta^{\rm
q.e.} (Ref.\protect\cite{fano})$
\\
\tableline
& 1/3     &  0.0779(10)        & 0.0825(12)       & 0.075(5)  & 0.0772  \\
& 1/5     &  0.0166(6)         & 0.0191(6)        &   ---     & 0.0173  \\
& 1/7     &  0.0063(4)         & 0.0070(5)        &   ---     &   ---   \\
\end{tabular}
\end{table}

\begin{table}
\caption{
Total gap for the fractional QHE with $\nu = 1/3$, 1/5, and 1/7. The
Monte Carlo results of Morf and Halperin \protect\cite{morfs} and
Bonesteel \protect\cite{bonesteel}, and extrapolated exact
diagonalization results of Fano {\it et al.} \protect\cite{fano} are
given for comparison.}
\begin{tabular}{ccccccc}
& $\nu$ & $\Delta_{\rm CF}$ & $\Delta_{\rm L}$ & $\Delta_{\rm L}
(Ref.\protect\cite{morfs})$ & $\Delta_{\rm CF} (\protect\cite{bonesteel})$ & $\Delta (Ref.\protect\cite{fano})$ \\ 
\tableline
& 1/3     &  0.1058(16)     & 0.1104(17)    & 0.092(4)  & 0.106(3)   & 0.1036(2)
\\
& 1/5     &  0.0258(9)      & 0.0283(9)     &  ---      & 0.025(3)   & 0.0244(3)
\\
& 1/7     &  0.0101(6)      & 0.0108(6)     &  ---      & 0.011(3)   &   ---    
\\
\end{tabular}
\end{table}

\newpage

\begin{figure}
\centerline{
\psfig{figure=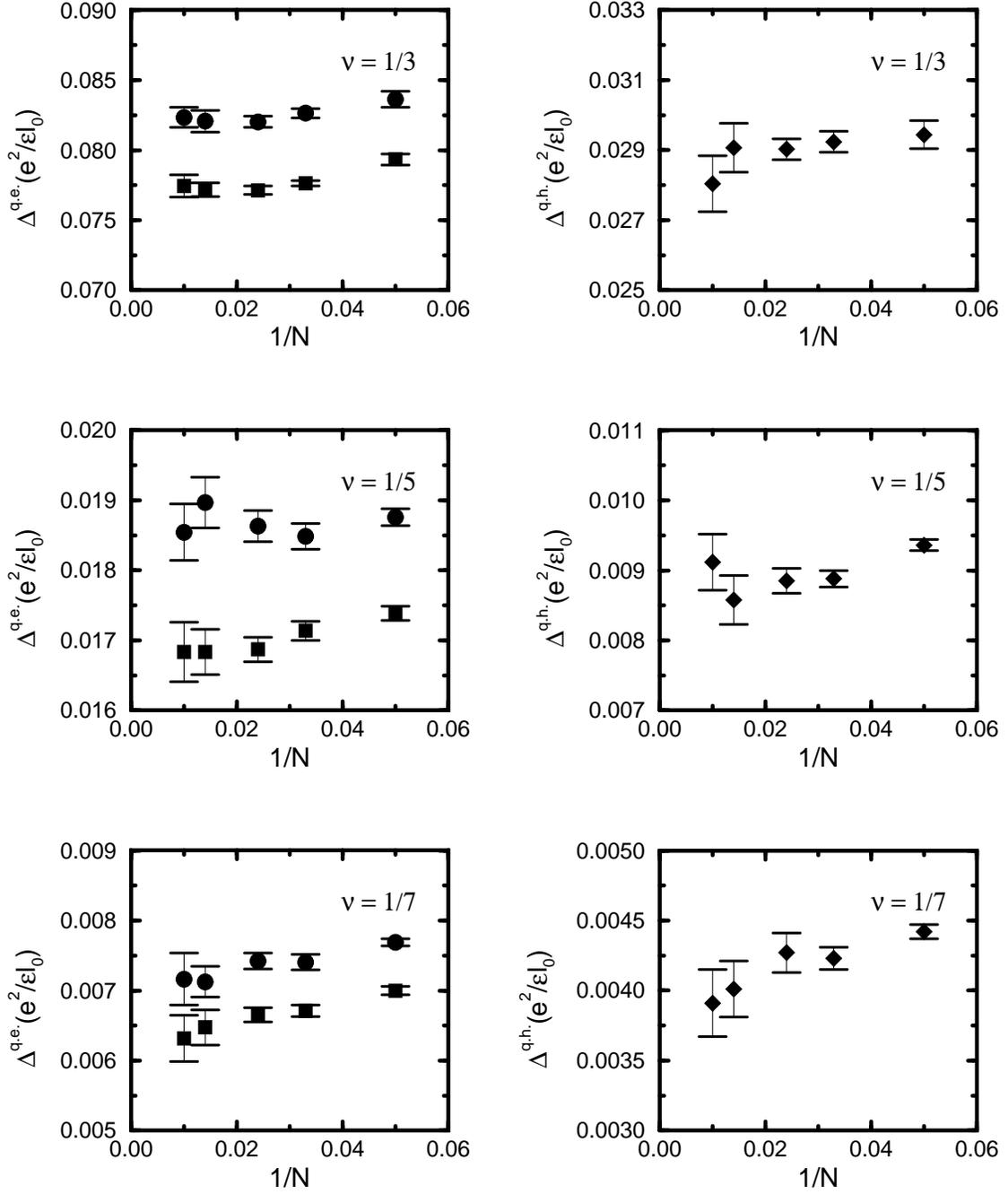,height=7in,angle=0}}
\
\
\caption{Proper energies for the Laughlin quasielectron (solid circles),
composite fermion quasielectron (solid squares), and quasihole (solid
diamonds) for filling fractions $\nu=1/3$, 1/5 and 1/7, plotted vs.
$1/N$.}
\end{figure}

\end{document}